# Predictive Structure-to-Thermal Conductivity Modeling Framework for BEOL Interconnect Stacks in Advanced Technology Nodes Enabled by Extensive Layer-Resolved Thermal Measurements

Zifeng Huang, Yiyang Sun, Tianyu Jia, Runsheng Wang, Zhe Cheng*
School of Integrated Circuits, Peking University, Beijing, China
Corresponding author: zhe.cheng@pku.edu.cn

***Abstract*—**The increasing structural complexity of BEOL interconnect stacks in advanced integrated circuits demands a structure-aware thermal conductivity ($\kappa$) modeling framework. However, generalizable models derived from layer-resolved thermal measurements that quantitatively capture the dependence of $\kappa$ on interconnect structures remain lacking, limiting predictive thermal analysis. Here, we establish an experimentally derived, structure-aware $\kappa$ modeling framework enabled by time-domain thermoreflectance (TDTR) measurements with ~100 nm depth resolution. Statistical analysis of a compiled dataset comprising over 40 experimentally measured layer-resolved $\kappa$ values across diverse BEOL layers reveals a generalizable empirical structure-to-$\kappa$ relationship, enabling predictive modeling based on interconnect structure. An intra-layer three-dimensional $\kappa$ distribution model based on effective medium theory and realistic layouts further resolves spatial $\kappa$ variations within practical interconnect layers. Together, these models establish an experimentally derived, structure-aware $\kappa$ modeling framework for predictive and generalizable thermal analysis of advanced interconnect stacks and 3D ICs.

## I. Introduction

Compared with conventional 2D chips, the BEOL interconnect stacks in three-dimensional integrated circuits (3D-ICs) introduce longer and more complex thermal transport paths between active layers and heat sinks. Meanwhile, the continuously increasing interconnect density further aggravates thermal accumulation, as shown in **Fig. 1** [1]. In advanced BEOL technologies, the thermal transport capability of interconnect layers is no longer determined solely by the intrinsic properties of bulk copper and dielectric materials, but is strongly governed by nanoscale structural features, including interconnect linewidth and the increasing contribution of Cu/dielectric interfaces. These structural effects lead to substantial variations in effective thermal conductivity ($\kappa$) among different interconnect layers, making conventional thermal models based on fixed material parameters insufficient for structure-aware prediction of heat dissipation behavior. Despite its critical role in thermal management and device reliability, experimentally established and generalizable models that quantitatively capture the dependence of $\kappa$ on BEOL structures remain lacking, hindering predictive thermal network modeling and hotspot analysis.

Here, we establish an experimentally derived, structure-aware $\kappa$ modeling framework for BEOL interconnect stacks by integrating layer-resolved TDTR measurements, effective medium theory (EMT), and realistic layout information. Statistical analysis of more than 40 layer-resolved $\kappa$ values across diverse BEOL layers reveals a generalizable empirical structure-to-$\kappa$ relationship, enabling predictive $\kappa$ modeling based on key interconnect parameters, including linewidth and Cu volume fraction. An EMT model further elucidates the role of Cu/dielectric composite structures in governing thermal transport and enables intra-layer spatial $\kappa$ mapping from realistic BEOL layouts, as shown in **Fig. 2** [2]. This framework provides a methodology for predictive thermal analysis of advanced ICs by capturing both layer-level $\kappa$ variations and intra-layer spatial distributions.

## II. Experimental Setup and Sample Preparation

The sample preparation and TDTR measurement procedures are illustrated in **Fig. 3**. The chip was fabricated using a commercial 22 nm process node. The five lowest interconnect layers (M1–M5, **Fig. 3(a)**) were sequentially exposed by plasma FIB milling, followed by Al transducer deposition for TDTR measurements (**Fig. 3(b)**). Thermal characterizations were performed from M1 to M5 to minimize the number of fitting parameters. The exposed surfaces exhibited sufficient smoothness and large-area uniform regions suitable for reliable TDTR measurements (**Figs. 3(c–d)**). TDTR is a pump–probe optical technique, and its experimental setup and measurement principles are schematically described in **Fig. 4** and Ref. [3].

The heat capacities used in the TDTR measurements are summarized in **Fig. 5**. As an example, the high-magnification SEM image of M5 in **Fig. 5(a)** was used to extract the in-plane Cu area fraction. By combining the SEM analysis with the TEM cross-sectional images shown in **Fig. 5(b)**, the Cu and dielectric volume fractions of each interconnect layer were determined. The corresponding effective volumetric heat capacities are summarized in **Fig. 5(c)**.

## III. Results and Discussion

**Figs. 6(a–b)** show the TDTR fitting curves for each interconnect layer measured using 5× and 20× objective lenses to exclude the spot-size dependence of the extracted $\kappa$ values. The extracted $\kappa$ are summarized in **Fig. 6(c)**, showing good consistency between the two measurements. **Fig. 6(d)** presents the sensitivity analysis of M5, confirming the reliability of the fitting results. Consistently low effective $\kappa$ approaching 1 W/m-K are obtained with both objective lenses, mainly attributed to the large Cu/dielectric thermal boundary resistance (TBR) induced by the high density of nanoscale interfaces and Ta barriers surrounding the Cu wires [4–6].

The physical origin of the thermal degradation was further elucidated by nanoscale structural and elemental analyses using BF-TEM and EDS, as shown in **Fig. 7**. **Fig. 7(a)** shows the confined sub-100 nm Cu interconnects, while **Fig. 7(b)** presents the Cu distribution. **Fig 7(c)** highlights the continuous Ta diffusion barrier layers surrounding the Cu wires, and **Fig. 7(d)** shows the corresponding Si and O elemental maps of the dielectric layer. The dense Ta barrier layers introduce additional interfaces, resulting in large Cu/dielectric TBR. Combined with grain boundary scattering, these interfacial effects significantly reduce the effective $\kappa$.

Based on the nanoscale structural analysis discussed above, the wire $\kappa$ was further evaluated following the framework shown in **Fig. 8 (a)**. Accordingly, the linewidth ($w$)-dependent wire $\kappa$ under different grain boundary (GB) conditions was calculated, as shown in **Fig. 8(b)**, together with DFT results from the literature [7]. By fixing the wire $\kappa$, the dielectric $\kappa$ and Cu/dielectric TBR were extracted through EMT fitting of the experimental data (**Fig. 8(c)**), and the extracted values are summarized in **Fig. 8(d)**. The upper bound of dielectric $\kappa$ ($\kappa_{\text{die}}$) was set as 1.5 W/m-K and $\kappa_{\text{die}}$ was assumed to be identical for layers M2–M5, which share a porous low-$k$ dielectric distinct from that of M1. The obtained TBR values are consistent with previously reported single-interface measurements [8].

While the EMT model reveals the physical origins of $\kappa$ degradation, a generalizable predictive relationship linking BEOL structural parameters to effective $\kappa$ is essential for extending thermal analysis beyond individual experimental measurements. As shown in **Fig. 9**, more than 40 layer-resolved TDTR measurement results were compiled and statistically analyzed to establish an empirical prediction model for the $\kappa$ of interconnect layers based on two key structural parameters: linewidth ($w$) and Cu volume fraction ($f$). The assumed formula has the form of:

$$\kappa_{\text{eff}} = \frac{(1-f)\kappa_{\text{die}} + f\kappa_{\text{Cu}}\dfrac{w}{w+A}}{1 + B\dfrac{f}{w}\left(1+\dfrac{w}{w_0}\right)^{-1}}$$

Here, the $\kappa_{\text{Cu}}$ and $\kappa_{\text{die}}$ are fixed to represent the intrinsic contributions of Cu and dielectric, with an electron mean free path $A$ of 39 nm to account for linewidth-dependent size effects. The fitting parameters $B$ and $w_0$ describe the strength and characteristic decay length of Cu/dielectric TBR, respectively, where $w_0$ represents the linewidth above which the interfacial effect becomes progressively less significant. **Fig. 9(a)** compares the model predictions using the best fit $B$ and $w_0$ (~120 nm) with experimental results, where the size of each data point represents the measured $\kappa$ value. The corresponding $\kappa$ contour map calculated from the model is also presented using the same color scale, showing good agreement between the model and experimental data. The dashed line indicates a linewidth of $3w_0$, corresponding to approximately 360 nm.

The model further reveals a critical interplay between $w$ and $f$. At ultra-narrow linewidths, increasing $f$ fails to improve the effective $\kappa$ and can even degrade it due to enhanced TBR penalties; a higher $f$ becomes beneficial only at larger linewidths (**Fig. 9(b)**). At a fixed $w$ = 50 nm (**Fig. 9(c)**), minimizing the interfacial parameter $B$ to 30% of its best-fit value yields a ~2.5-fold increase in the $\kappa_{\text{eff}}$, highlighting the dominant role of interfacial resistance.

While the empirical model captures layer-level $\kappa$ variations, the local structural information from realistic layouts enables further reconstruction of intra-layer $\kappa$ distributions. Specifically, Cu pixels were extracted from the layout, which was then discretized into Cu- and dielectric-containing cells. The local structural parameters of each cell were used as inputs for the EMT model, with the parameters extracted in **Fig. 7**, to calculate the corresponding $\kappa$ values of each cell. The reconstruction process is illustrated in **Fig. 10** with a cell size of 500 nm. Layer-resolved $\kappa$ mapping was further performed using TDTR with different objective lenses and spatial step sizes to validate the calculated results, as shown in **Fig. 11(a–b)**. The obtained $\kappa$ profiles enable a direct comparison between experimental measurements and model predictions. As shown in the enlarged view in **Fig. 11(c)**, the reconstructed $\kappa$ distributions agree well with the TDTR mapping results, accurately capturing the low-$\kappa$ regions and reproducing the measured spacing between adjacent low-$\kappa$ strips (~20 µm).

## IV. Finite Element Simulation

The experimentally calibrated $\kappa$ distributions were further incorporated into finite element analysis (FEA) model enabled by COMSOL to evaluate their impact on thermal prediction, as shown in **Fig. 12**. The reconstructed $\kappa$ maps and estimated heat-source distributions (blue regions) with corresponding spatial resolutions were implemented in the simulations (**Figs. 12(a–b)**). Identical total power dissipation was maintained for all simulation cases. The resulting maximum temperature rise (**Fig. 12(c)**) increases as finer $\kappa$ and heat-source distributions are resolved, because localized low-$\kappa$ regions and heat generation are progressively captured [9]. This trend indicates that conventional thermal models based on averaged material properties may underestimate hotspot temperatures by neglecting local thermal inhomogeneity. These results further demonstrate the importance of incorporating realistic $\kappa$ distributions for thermal analysis of BEOL interconnects.

## V. Conclusion

In summary, we establish a structure-aware $\kappa$ modeling framework for BEOL thermal analysis by deriving a generalizable empirical structure-to-$\kappa$ relationship from extensive TDTR measurements and reconstructing layer-resolved $\kappa$ distributions from realistic layouts. This framework enables high-resolution prediction of thermal variations and provides a quantitative methodology for thermal analysis and management in advanced ICs. Compared with conventional simulation-based approaches, this framework enables efficient $\kappa$ prediction through a simplified $\kappa$ extraction methodology while maintaining physical relevance (**Fig. 13**). More importantly, the proposed modeling framework can be readily extended to scaled BEOL technologies, providing a general methodology for thermal analysis of future technology nodes.

## Acknowledgment

The authors acknowledge the financial support from the National Key Research and Development Program of China (Grant No. 2024YFA1207901) and the National Natural Science Foundation of China (NSFC) (Grant Nos. 62574007, T2550270).

## References

[1] W.-Y. Woon *et al.*, *Nat Rev Electr Eng*, vol. 2, no. 9, pp. 598–613, 2025. [2] C.-W. Nan *et al.*, *J. Appl. Phys.*, vol. 81, no. 10, pp. 6692–6699, 1997. [3] D. G. Cahill, *Rev. Sci. Instrum.*, vol. 75, no. 12, pp. 5119–5122, 2004. [4] T. Zhan *et al.*, *ACS Appl. Mater. Interfaces*, vol. 14, no. 5, pp. 7392–7404, 2022. [5] T. Zhan *et al.*, *ACS Appl. Mater. Interfaces,* vol. 12, no. 19, pp. 22347–22356, 2020. [6] Y. He *et al.*, *EDTM*, 2026. [7] A. Wang *et al.*, *Appl. Phys. Lett.*, vol. 124, no. 21, pp. 212202, 2024. [8] Z. Huang *et al.*, *IEDM*, 2025. [9] B. Vermeersch *et al.*, *IEDM*, 2024. [10] T. Chou *et al.*, *IEDM*, 2025. [11] L. Wang *et al.*, *IEDM*, 2024. [12] S. Mishra *et al.*, *VLSI*, 2024.

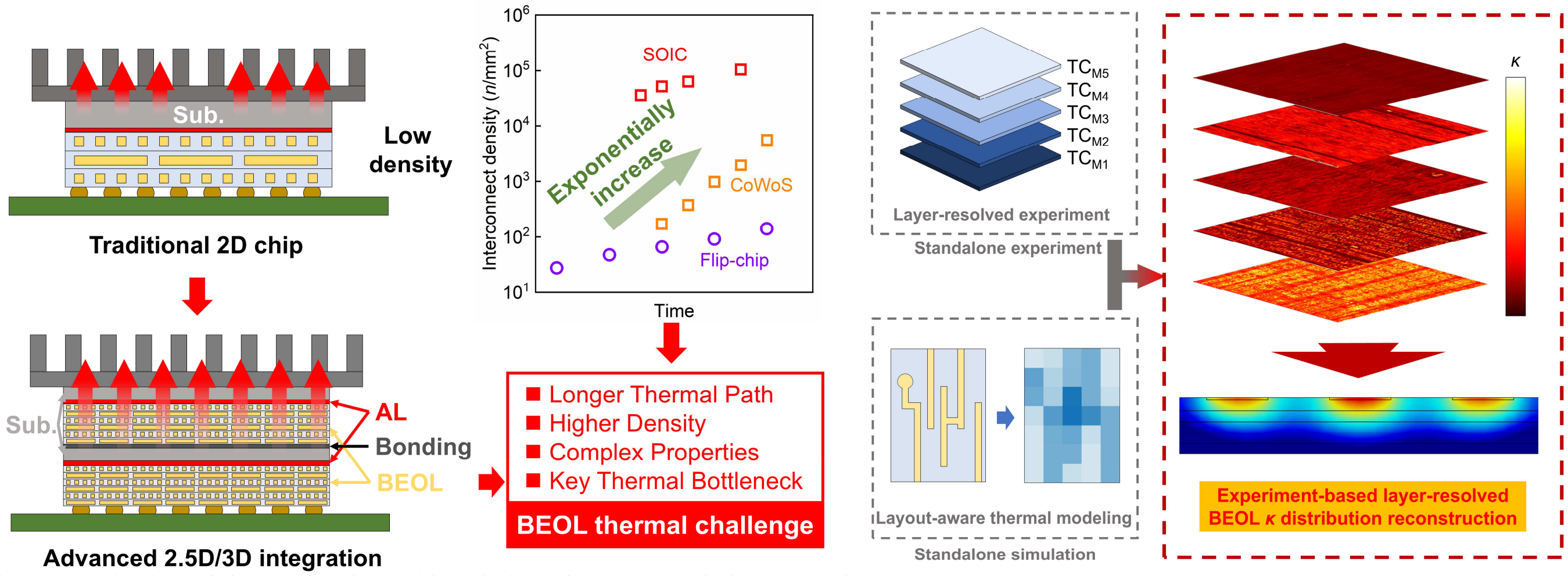

Fig. 1. Motivation of this work: The rapid evolution of conventional chips toward advanced 2.5D/3D integrated systems has led to an exponential increase in interconnect density. This increase results in longer thermal dissipation paths and more complex heat transport characteristics, making the BEOL interconnect stack a major bottleneck for advanced thermal management.

Fig. 2. Comparison between conventional standalone experimental/simulation approaches and the proposed coupled experiment–simulation framework for layer-resolved BEOL thermal conductivity reconstruction.

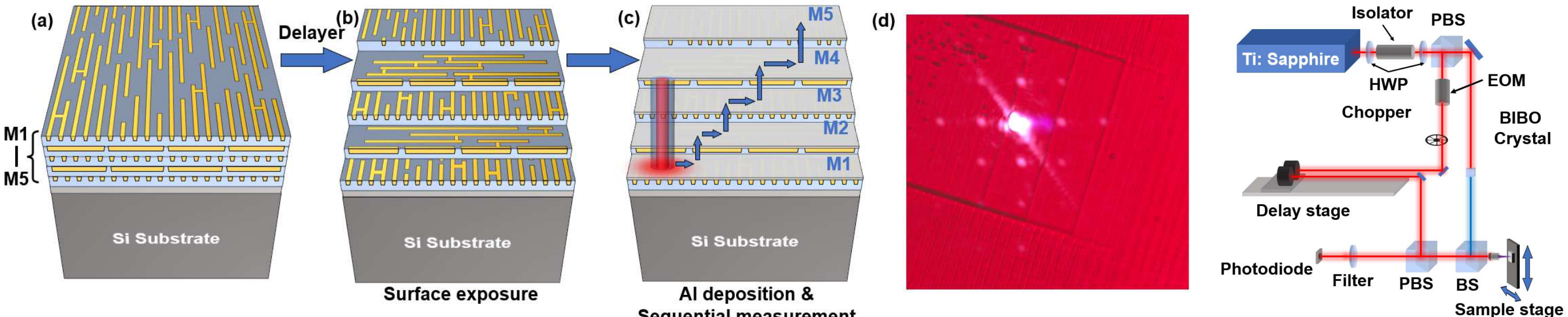

Fig. 3. Schematic of sample preparation and the measurement workflow. (a) Original BEOL structure of the chip. (b) Exposed BEOL stack after P-FIB delayering. (c) Al transducer layer deposition followed by sequential measurements. (d) Optical and surface morphology observation during the actual experiment.

Fig. 4. Schematic illustration of the TDTR experimental setup used in this study.

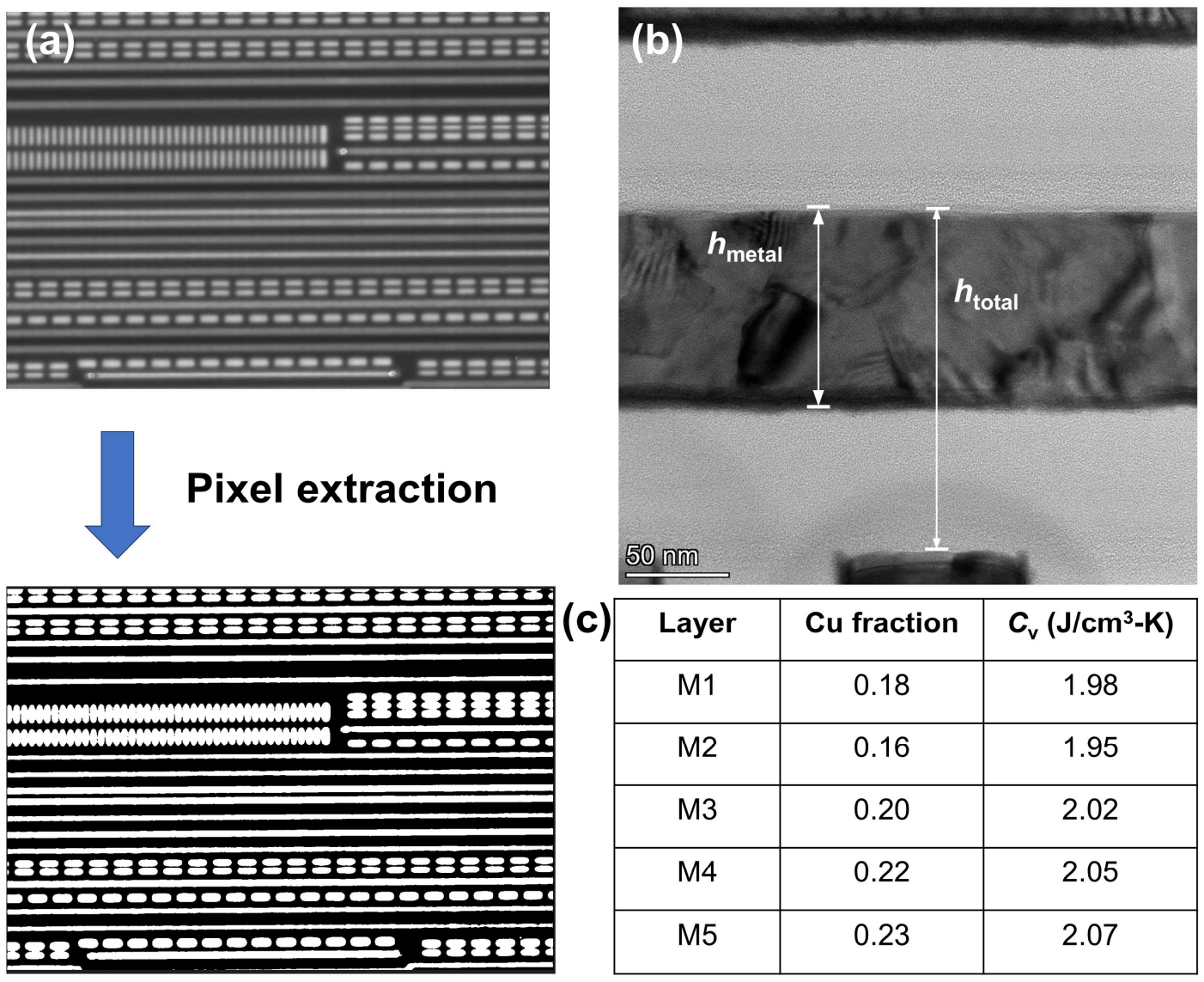

| Layer | Cu fraction | $C_v$ (J/cm³-K) |
|---|---|---|
| M1 | 0.18 | 1.98 |
| M2 | 0.16 | 1.95 |
| M3 | 0.20 | 2.02 |
| M4 | 0.22 | 2.05 |
| M5 | 0.23 | 2.07 |

Fig. 5. Workflow of volumetric heat capacity extraction for individual layers: (a) Identification of Cu regions by extracting high-brightness pixels from the SEM image to yield the surface Cu fraction. (b) Determination of the total Cu fraction by combining $h_{\mathrm{metal}}$ and $h_{\mathrm{total}}$. (c) Extracted volumetric heat capacity ($C_{\mathrm{V}}$) of each individual interconnect layer.

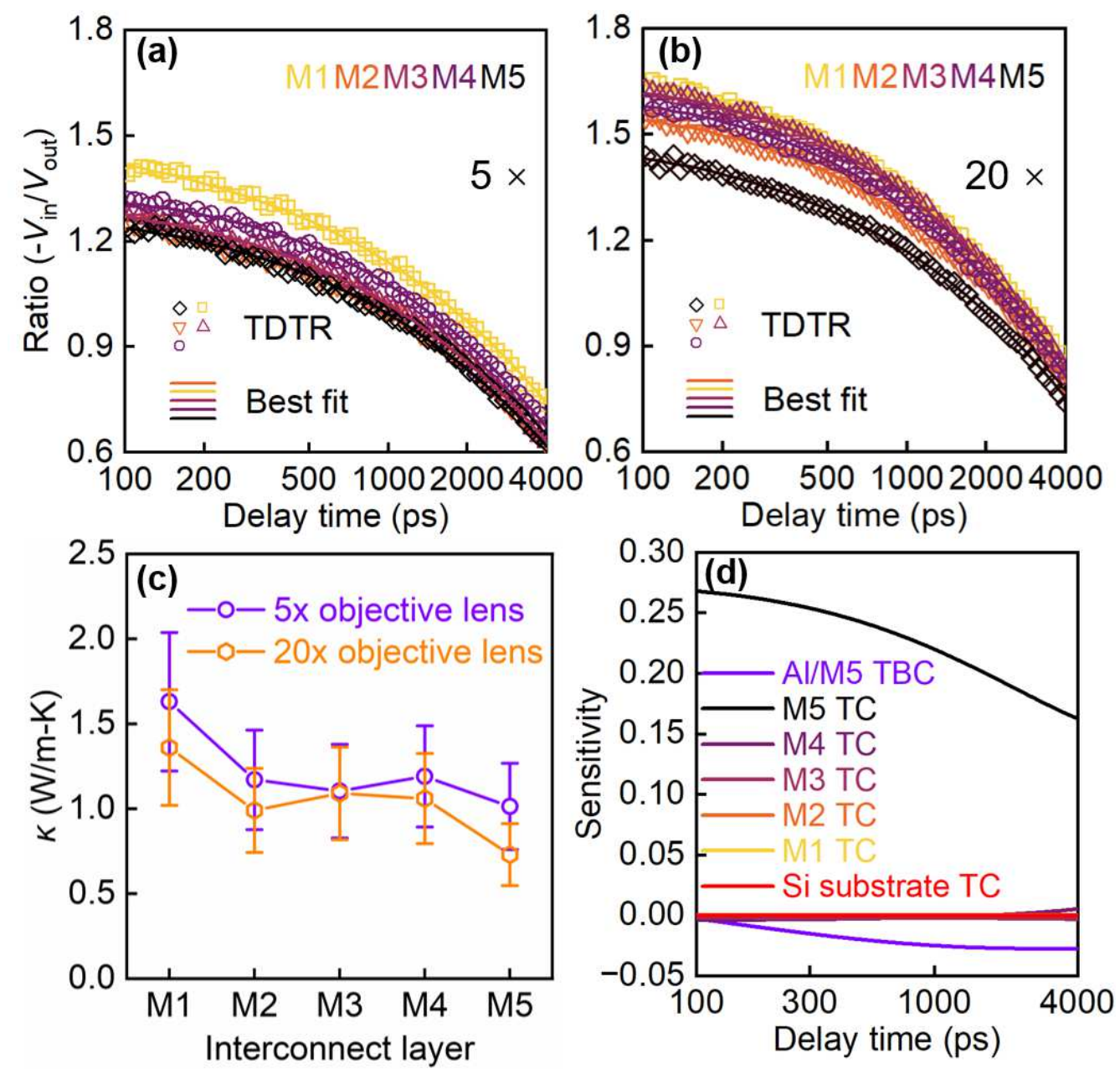

Fig. 6. TDTR data analysis and fitting: (a)-(b) Measured TDTR ratio signals and corresponding best fits for individual layers with 5× and 20× objective lenses. (c) Measured $\kappa$ of individual layers compiled using both 5× and 20× objective lenses. (d) Sensitivity analysis curves for the M5 layer measurement.

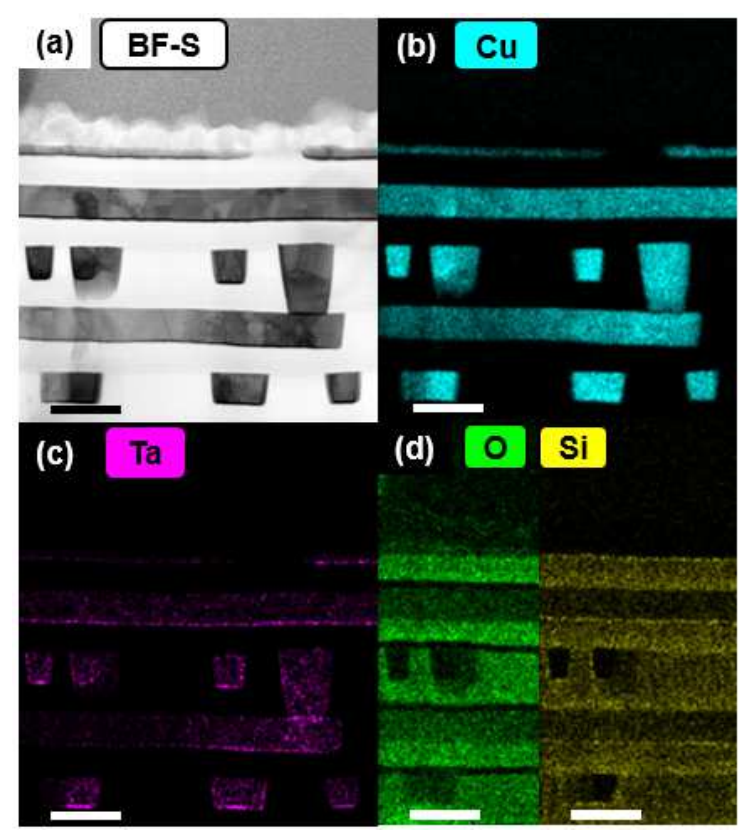

Fig. 7. Structural and elemental analysis of the BEOL stack: (a) BF TEM image of the interconnect stack; (b)-(c) EDS map showing the Cu and Ta distribution; (d) EDS maps of Si and O in the dielectric. The scale bar is 200 nm.

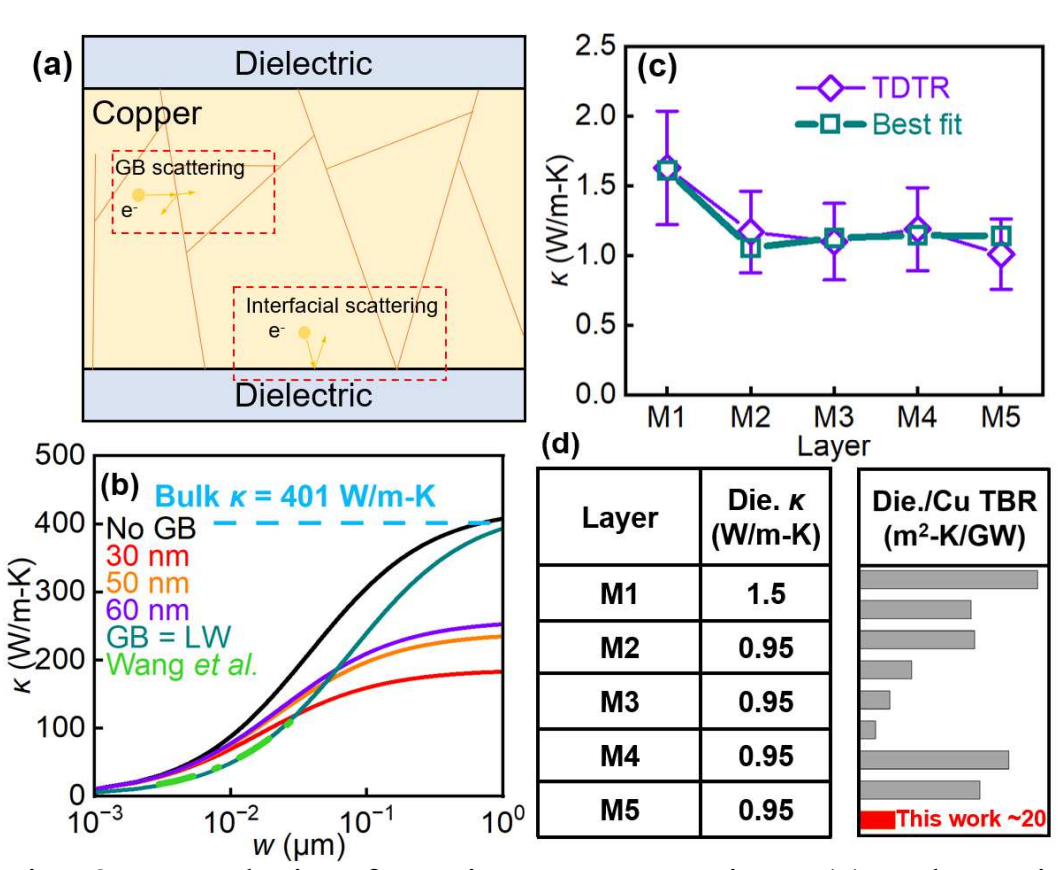

Fig. 8. $\kappa$ analysis of Cu interconnect wires: (a) Schematic illustration of the two dominant scattering mechanisms in the wire. (b) Calculated thermal conductivity of Cu wires as a function of $w$. (c) Best-fit results of the effective $\kappa$ model to the experimental data. (d) Thermal transport parameters extracted from the fitting procedure and their comparison with different Cu/dielectric TBRs from [8].

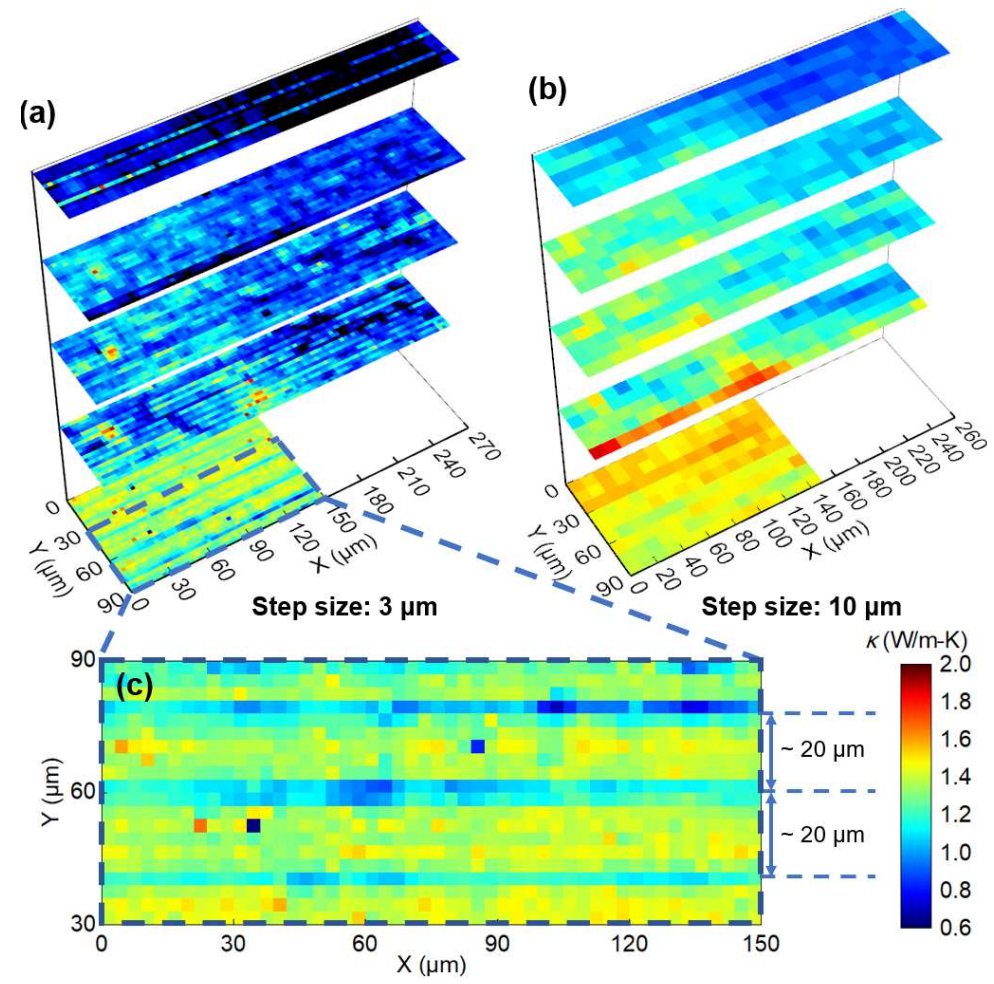

Fig. 11. Layer-resolved $\kappa$ mapping by TDTR: (a) Mapping results obtained using a 20× objective lens with a step size of 3 μm, and (b) Mapping results obtained using a 5× objective lens with a step size of 10 μm. (c) Detailed $\kappa$ map of M1, with the spatial intervals between adjacent strips highlighted.

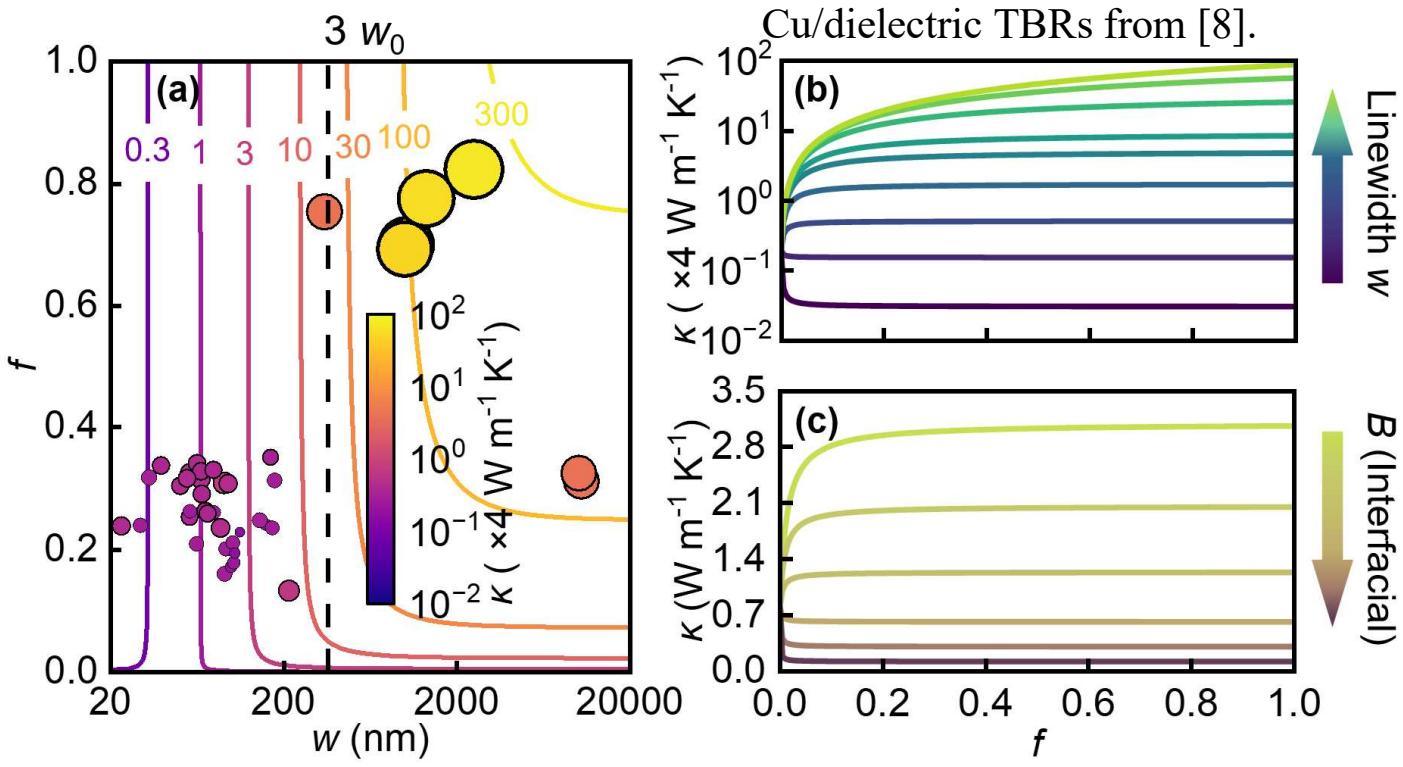

Fig. 9. Empirical $\kappa$ prediction model of BEOL interconnects as a function of $w$ and $f$. (a) Contour map of the fitted effective $\kappa$ as a function of $w$ and $f$ empirical model. Colored circles represent experimentally measured $\kappa$ values. (b) Evolution of the effective $\kappa$ with $w$ for different $f$, (c) Dependence of the effective $\kappa$ on $f$ for different interfacial parameter $B$, showing that stronger interfacial effect (larger $B$) increasingly suppresses the thermal conductivity.

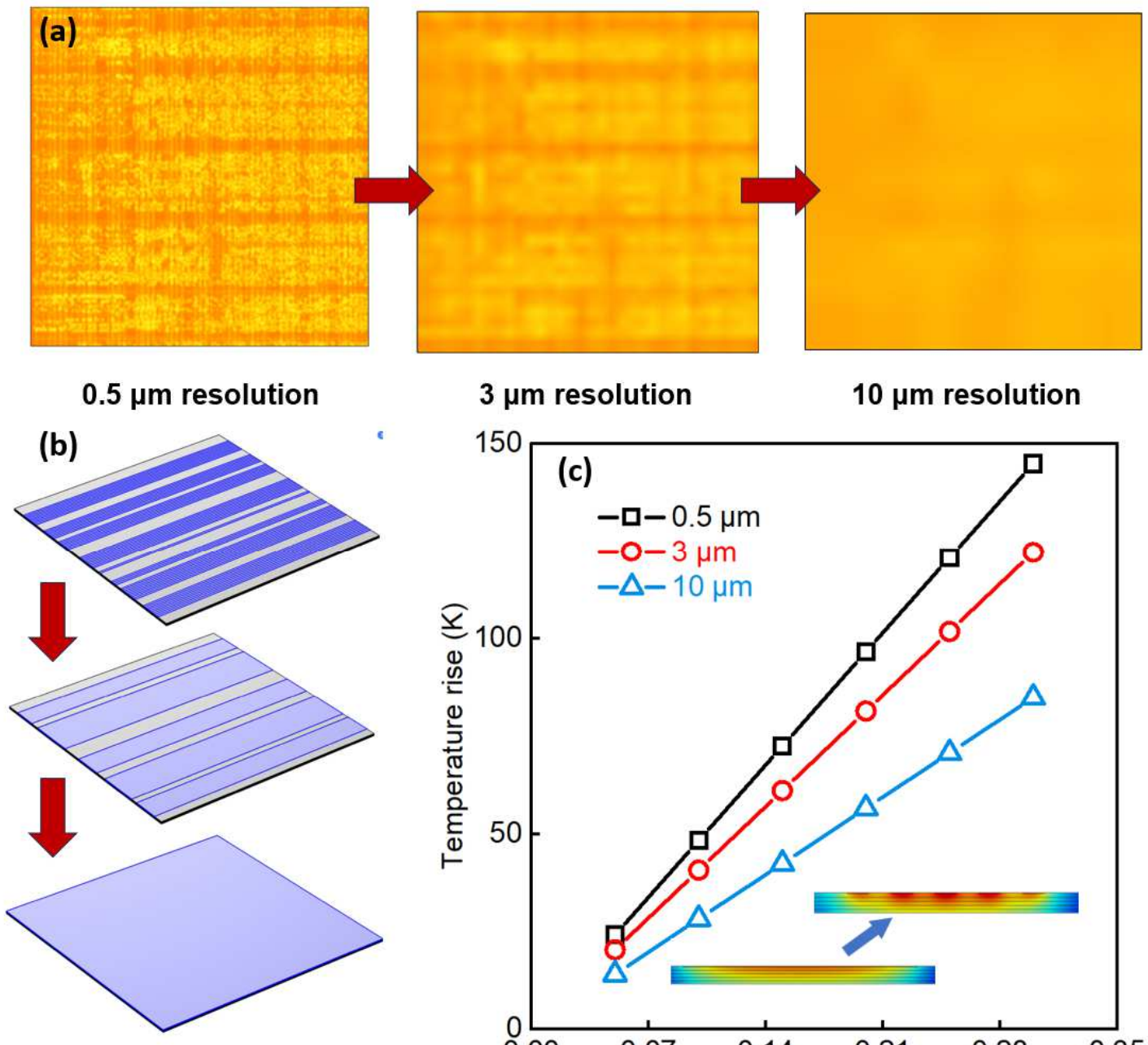

Fig. 12. FEA incorporating detailed $\kappa$ distributions: (a) $\kappa$ distribution maps at different spatial resolutions, with M1 shown as an example. (b) Corresponding FEA model with resolution-dependent heat source (blue regions) definitions. (c) Simulated temperature rise for models with different spatial resolutions. The inset illustrates the evolution of the temperature profiles as the spatial resolution is refined from 3 μm to 500 nm.

Fig. 10. Effective $\kappa$ map extracted from layout information: (a) Schematic illustration of the procedure used to derive the effective $\kappa$ map from the original layout image, and (b) The resulting effective $\kappa$ maps extracted from the layout figures of metal layers M2–M5. The scale bar is 40 μm.

| | Modeling foundation | Generalizable relationship | Layout-aware $\kappa$ reconstruction | $\kappa$ calculation method |
|---|---|---|---|---|
| This work | **Extensive experiment** | √ | √ | **Exp.+ Formula** |
| IEDM 2025 [8] | Experiment | - | - | Exp. |
| IEDM 2025 [10] | Simulation | - | Partial (M0) | Hybrid FEA* |
| IEDM 2024 [11] | Simulation | - | - | FEA |
| IEDM 2024 [9] | Simulation | - | - | FEA+Fixed |
| VLSI 2024 [12] | Simulation | - | - | Fixed |

Fig. 13. Comparison of the proposed framework with previous BEOL thermal modeling approaches in terms of modeling foundation, generalizable structure–$\kappa$ relationships, layout-aware $\kappa$ reconstruction, and $\kappa$ evaluation methods. *Note: For Ref. [10], a detailed model was constructed for the M0 layer, while the remaining layers were treated as an effective $\kappa$ obtained from an additional FEA model.